\documentstyle{l-aa}
\begin{document}

\thesaurus{03(11.01.2; 11.02.1;  13.18.1; 13.25.2)}

\def\nh{$N_{\rm H}$\ } 
\def\vv{V\'eron-Cetty  \& V\'eron }
\def\arx{$\alpha_{\rm {rx}}\,$}
\def\aro{$\alpha_{\rm {ro}}\,$}
\def\aox{$\alpha_{\rm {ox}}\,$}
\def\ax{$\alpha_{\rm {x}}\,$}
\def\am{$\langle\alpha\rangle\,$} 
\def\del{$\delta_{\rm IC}\,$}
\def\ch{$C_{\rm{H}}\,$}
\def\cs{$C_{\rm{S}}\,$}

\input psfig.tex  
               
\renewcommand{\textfraction}{0.0}

\title{ROSAT Observations of BL Lacertae Objects}
\author{G. Lamer\inst{1} 
\and H. Brunner\inst{1,2}  
\and R. Staubert\inst{1} }

\offprints{G. Lamer}

\institute{Institut f\"ur Astronomie und Astrophysik, Abt. Astronomie,
           Universit\"at T\"ubingen, 
           Waldh\"auserstr. 64, 
           D-72076 T\"ubingen, Germany
\and       Astrophysikalisches Institut Potsdam,
           An der Sternwarte 16,
           D-14482 Potsdam, Germany}

\date{Received  ; accepted }

\maketitle

\begin{abstract}
We present soft X-ray spectra of 74 BL Lacertae objects observed 
with the {~PSPC} detector on board of the {~ROSAT} satellite.
The sample contains all BL Lac objects detected during the 
pointed observation phase as a target or serendipitously.
We have investigated the soft X-ray and broad band spectral properties
and discuss the consequences for the X-ray emission processes.
For the first time a clear dependence of the  X-ray spectral steepness
on the radio to X-ray spectral energy distribution is found:
 \arx and \ax are {\em correlated} in the X-ray selected  (XBL) subsample and 
{\em anticorrelated} in the radio selected (RBL) subsample.
The objects with intermediate
\arx values thus do have the steepest soft X-ray spectra.
Simulated {~PSPC} spectra based on a set of simple two component 
multifrequency spectra are in good agreement with the 
measurements and suggest a broad range of synchrotron cutoff energies.
 
We have calculated synchrotron self-Compton beaming factors for a subsample
of radio bright objects and find a correlation of the beaming factors 
\del with \arx and \ax.  
The most extreme RBL objects are very similar to flat spectrum radio
quasars in all their broad band and X-ray properties.

 
\end{abstract} 

\section{Introduction}

BL Lac objects are active galactic nuclei which
by definition  do not have strong emission lines, are highly variable, 
and show strong polarization in the radio to optical emission.
Commonly the properties of BL Lacs are explained by the concept
that the emission in all spectral bands is dominated by  relativistic
jets. Relativistic electrons emit synchrotron radiation and 
may scatter up either synchrotron photons (synchrotron self-Compton
emission) or photons from other regions (e.g. from an accretion disc) to 
higher energies. 
   
Originally most BL Lacs have been found as counterparts of flat spectrum
radio sources (radio selected BL Lacs, RBLs), but an increasing number of 
less radio bright objects were discovered by the identification of 
X-ray sources (XBLs). Due to the lack of spectral features in the 
optical no complete optically selected samples exist. The total 
number  of BL Lac objects in the catalogue of \vv (\cite{veron}) is 
less than 200. A new catalogue consisting of 233 sources is being
published by Padovani \& Giommi (\cite{PG95b}). 
The ongoing search for new BL Lacs from {~ROSAT} sources 
(Kock et al. \cite{kock}, Nass et al. \cite{nass}) will 
significantly increase the number of known objects.
   
Recently BL Lacs have gained great interest due to the detection of 
several objects in  high energy $\gamma$-rays by EGRET on the Compton
Gamma Ray Observatory (von Montigny et al. \cite{montigny}).    

There is some evidence that RBLs and XBLs form distinct subclasses
as they show a bimodal distribution in the plane of the broad band
spectral indices \aro versus \aox (e.g. Giommi
et al. \cite{giommi}). 
Regardless of the discovery waveband we will use these terms for
a distinction of the subclasses based on the spectral energy 
distribution: 
RBL is used for {\it radio} bright objects 
having \arx$>$0.75 and XBL for {\it X-ray} bright objects with \arx$<$0.75.

There have been several attempts to explain the physical differences 
of the XBL and RBL objects and their relation to flat spectrum radio quasars
({~FSRQs}). The concept that parts of the continuum emission
from radio loud AGN arises from jets of radiogalaxies more or 
less aligned with the line of sight is widely accepted.  

Based on the study of number count relations and  luminosity functions
several authors (e.g. Padovani \& 
Urry \cite{PU}) proposed that BL Lacs are the beamed subpopulation of FR I 
galaxies with RBLs having higher beaming factors than XBLs. 
Ghisellini \& Maraschi (\cite{GM}) discussed an ``accelerating jet'' model with 
lower bulk Lorentz factors $\Gamma$ in the X-ray  emitting regions resulting
in broader beaming cones for the X-ray emission and narrow radio cones.
Celotti et al. (\cite{celotti}) developed a ``wide jet'' model with 
geometrically wider  opening angles in the inner, X-ray emitting, parts of
the jet. Assuming that RBLs have smaller viewing angles
than XBLs, both models are able to explain both the relative numbers and 
different spectral energy distributions of XBLs and RBLs with an intrinsically
uniform population of objects. Maraschi \& Rovetti (\cite{maraschi})
have extended these
considerations on {~FSRQs} and propose that all radio loud AGN only essentially 
differ in viewing angle and intrinsic power of the central engine.        

An alternative approach to explain the differences between XBLs and RBLs 
was made by Padovani \& Giommi  (\cite{PG95a}) with a ``different energy 
cutoff'' hypothesis. They argue that both types
form a uniform class of objects spanning a wide range in the intrinsic
energy distribution caused by  different cutoff frequencies of the
synchrotron component.      

The  X-ray spectra of RBLs in average were found
to be significantly steeper than the  spectra of (higher redshifted) {~FSRQs} in 
{\it Einstein} IPC (Worrall \& Wilkes \cite{ww}) and
{~ROSAT} {~PSPC} (Brunner et al. \cite{brunner}) investigations.
Furthermore, the X-ray spectral indices of BL Lacs showed a broad 
distribution in both investigations.
The mean X-ray spectra of XBLs and RBLs were not found to be 
significantly different in  {\it Einstein} IPC (Worrall \& Wilkes \cite{ww}),
EXOSAT ME + LE (Sambruna et al. \cite{sambruna}), and {~ROSAT} {~PSPC} 
(Lamer et al. \cite{lamer}) observations.
Ciliegi et al. (\cite{ciliegi}) found significant steepening of the 
mean spectrum of XBLs between the soft (0.2--4 keV) and medium (2--10 keV)
energy X-ray band.

The {~ROSAT} data archive presently comprises
the largest X-ray database for BL Lac objects
collected by a single instrument.
In this paper we present the analysis of X-ray and broad band 
spectra of 74 BL Lac objects with the X-ray data obtained 
from pointed {~ROSAT} observations.  
We find a strong interdependence of the X-ray spectral index \ax with
the radio to X-ray energy index \arx which we interprete as the
signature of two spectral components intersecting each other at 
different frequencies.

\section{The sample}

Due to the difficulties in the identification and classification 
of BL Lacs only relatively few objects form complete flux limited 
samples.
Nearly complete samples were
selected from radio sources (1~Jy sample, Stickel et al. \cite{stickel}; 
34 objects)   
and with limited sky coverage from soft X-ray sources 
(EMSS sample, Morris et al. \cite{morris}; 22 objects).
Our sample comprises all BL Lacs listed in the catalogue of 
V\'{e}ron-Cetty \& V\'{e}ron (\cite{veron}) and of which {~ROSAT} {~PSPC} 
observations exist in the archives. Only sources which had been detected
with more than 50 net counts have been analysed: 74 objects in total.
In case of multiple {~ROSAT} observations of an object the longest
observation available at the time of analysis has been selected.
Objects in the catalogue which have meanwhile been classified as quasars
or as radio galaxies have not been included  
(e.g. Stickel et al. \cite{stickel2}). 
The resulting coverage of various complete samples of BL Lac objects 
is listed in Table \ref{samples}. According to their radio to X-ray 
energy distribution 40 objects have been assigned to the XBL subsample,
34 objects to the RBL subsample (see Sect. 3).

\begin{table}
\begin{flushleft}
\caption[]{\label{samples}Coverage of complete samples}
\begin{tabular}{lll}
\hline\noalign{\smallskip}
sample       & Reference           & observed (total) \\
\noalign{\smallskip}
 \hline\noalign{\smallskip}
1~Jy (5 GHz) & Stickel et al. (\cite{stickel}) & 29 (34)          \\
EMSS         & Morris et al.  (\cite{morris}) & 19 (22)          \\
S5 ($\delta>70^{\circ})$&Eckart et al. (\cite{eckart})&  5 (5)   \\
\noalign{\smallskip}
\hline
\end{tabular}
\end{flushleft}
\end{table}

Throughout this paper the catalogue designations according to \vv 
(\cite{veron}) are used.      

\section{ROSAT observations and data analysis}

Archival data were taken from the {~ROSAT} data archives 
at MPE (Garching) and at GSFC (Greenbelt).
Both the author's proprietary data and archival data were reduced in
the same way using the EXSAS software (Zimmermann et al. \cite{zimm}).
Table~\ref{obs} list the objects, {~ROSAT} observation request (ROR) numbers, and
dates of observations which have been analysed.   

\begin{table*}
\caption[]{\label{obs}List of {~PSPC} observations 
\hspace{4.5cm} {\bf Table 2.} -- Continued}
\par{\psfig{figure=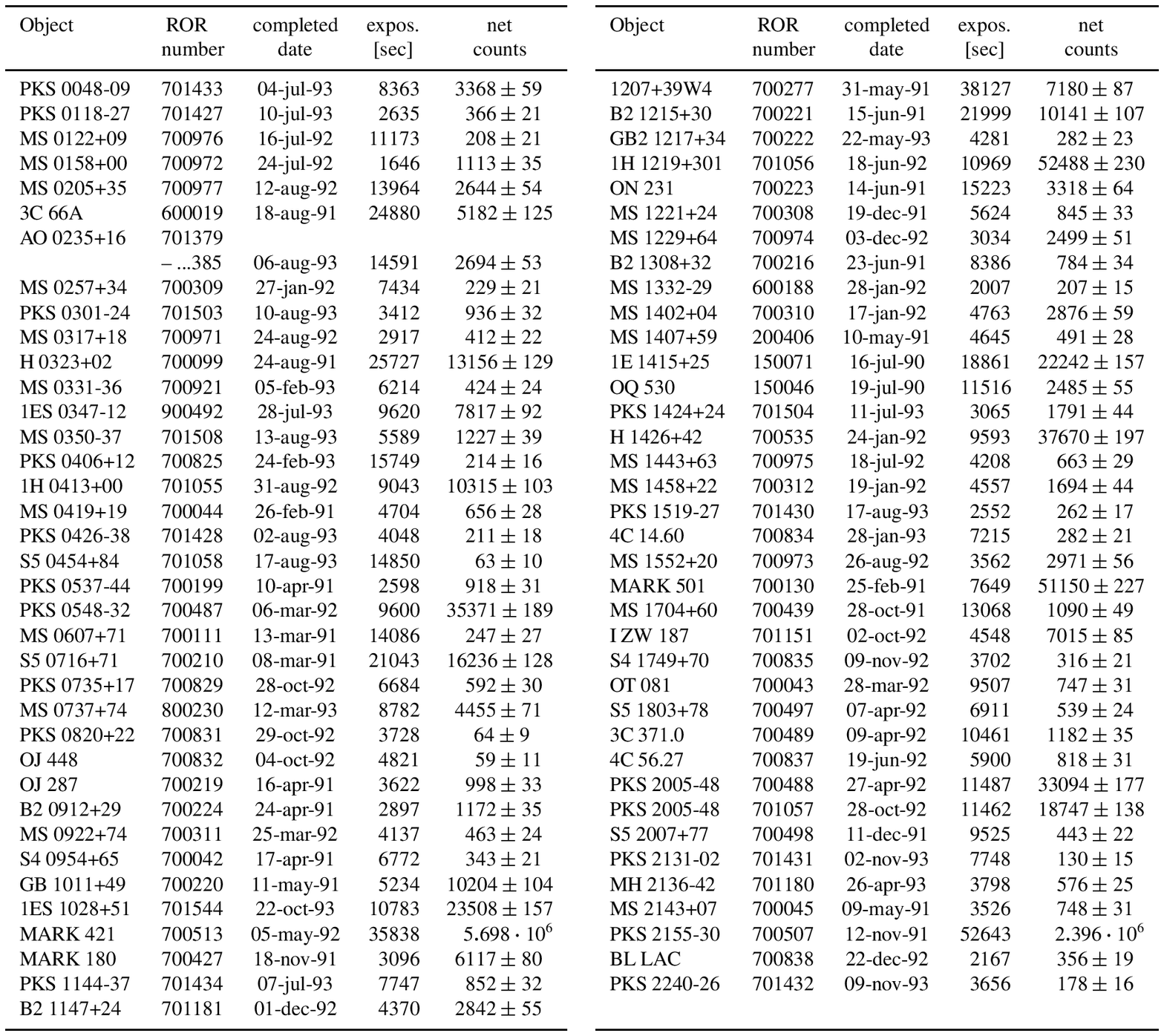,width=18truecm}}
\end{table*}

The source photons were extracted 
within a circle of radius 100''--200'' (depending on the signal to noise 
ratio) and the background determined in an annulus of radii 250'' and 
500''.  We produced spectra in the energy range 0.1--2.4 keV of all objects
by binning according to pulse height amplitude, yielding  a SNR per
spectral bin ranging from 
4 for the weakest sources and 50 for the strongest.  
All spectra were background subtracted and corrected for telescope
vignetting, dead time losses, and incomplete extraction of source photons
with respect to the point spread function.
The pulse height spectra were then fitted by power law spectra 
combined with the absorption model of Morrison and McCammon (\cite{mmcc}
).
For sources  with more than 250  detected counts
both fits with an absorbing column density \nh fixed to the galactic 
value (Stark et al. \cite{stark}, Elvis et al. \cite{elvis}) and free \nh were
performed.  For the weaker sources  only fits with fixed \nh were
obtained.

In general the latest version of the {~PSPC} detector response 
matrix (nr. 36) has been used for the fits. Except for observations carried
out before fall 1991, when the gain setting of the {~PSPC} was different,
an earlier version (nr. 6) was used.  

We combined the {~ROSAT} measurements with noncontemperaneous flux 
measurements at 5~GHz and in the optical V band (both taken from \vv)
in order to calculate the broad band spectral indices \arx, \aro,
and \aox. Sources with  \arx $> 0.75$ were assigned
to the RBL subsample and sources with \arx $<0.75$ to the 
XBL subsample.      

Power law spectral indices $\alpha$ are given as energy indices 
($f_{\nu}\propto \nu^{-\alpha}$) throughout the paper.  

We  used a maximum likelihood (ML)  method to deconvolve the measurement errors
and the intrinsic distribution of the X-ray spectral indices and other 
measured parameters when calculating mean values and their errors (see 
Worrall \cite{worrall} for a description of the method and Brunner et al. 
\cite{brunner} as a recent application). 
Assuming that both the intrinsic distribution of a parameter $p$
and the distribution of measurement errors are Gaussian, 
confidence contours of the mean $\langle p \rangle$ 
and width $\sigma_{\rm G}$ of the intrinsic distribution of $p$ can be
calculated.

\section{Results of spectral analysis}

\begin{table*}
\par{\psfig{figure=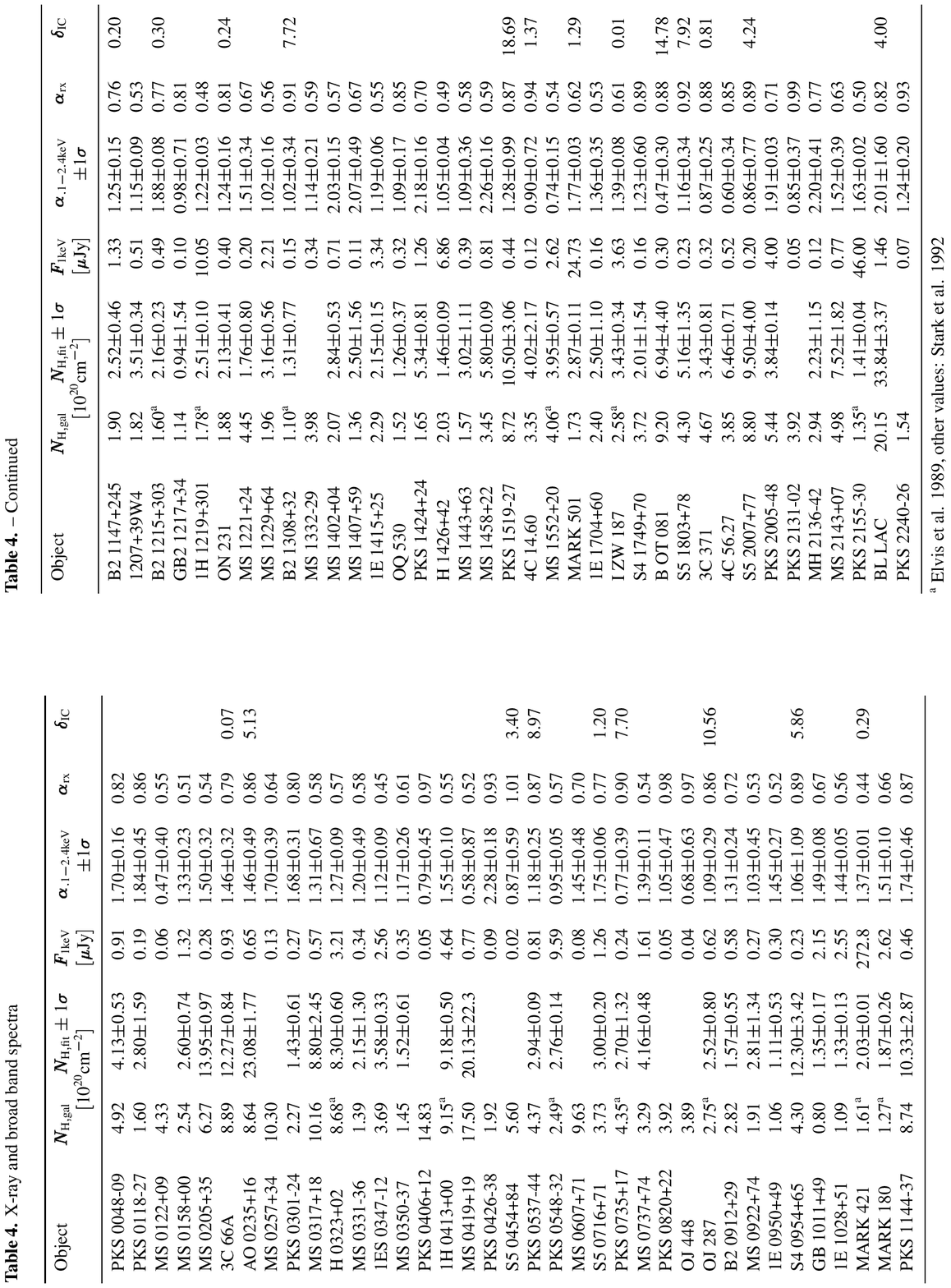,width=18 truecm}}
\end{table*}
\stepcounter{table}

\begin{figure}[htb]
\par\centerline{\psfig{figure=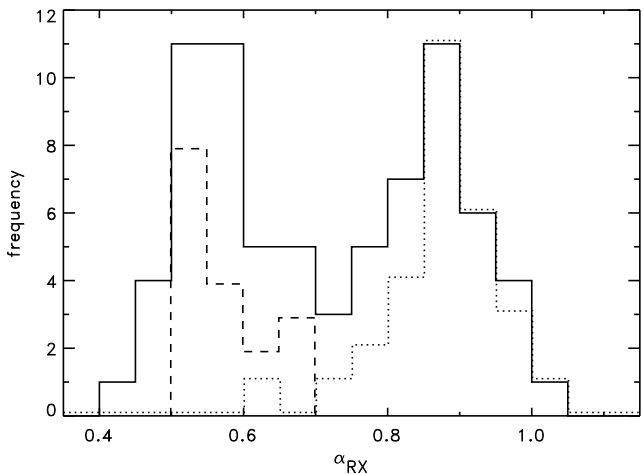,width=8.8 truecm}}
\caption[]{\label{arx}Distribution of \arx in the full sample (solid line),
in the EMSS sample (dashed), and in the 1Jy sample (dotted line).}
\end{figure}

The distribution of \arx (Fig.~\ref{arx}) is double peaked with a gap
at \arx$=0.6$--0.8. As can been seen from Fig.~\ref{arx}, all EMSS objects
exhibit an XBL energy distribution. Two objects of the 1~Jy sample
have \arx$<0.75$, they therefore belong to the XBL sample.  This confirms,
based on homogeneus X-ray data,  
the relations given by Padovani \& Giommi (\cite{PG95a}).

Figure \ref{ro} shows the interdependence of the calculated broad band 
spectral indices $\alpha_{\rm ox}$ and $\alpha_{\rm ro}$. Note that the gap near 
\arx = 0.75 is dominated by objects which do neither belong to the 
EMSS sample nor to the 1~Jy sample. This indicates that the gap is due
to selection effects and may be
filled in the future when radio or X-ray selected objects at lower
 flux limits will be identified.

Single power law  spectra with photoelectric absorption due to the 
interstellar medium in general yielded acceptable fits if the absorbing
column density \nh was left free to vary.
The results of the spectral fits with free \nh and the broad band 
spectral indices are given in Table 3. If a source was detected
with less than 250 counts, the entry $N_{\rm H,fit}$ is
omitted and the results with fixed \nh are given.    
In order to investigate whether deviations of the resulting 
\nh values from galactic HI radio measurements
are significant we calculated the difference $\Delta$\nh for
each object. The  error of the difference was calculated
by quadratic addition of the X-ray and radio measurement errors using 
$10^{20}$cm$^{-2}$ for the Stark et al. (\cite{stark}) values and 
$10^{19}$cm$^{-2}$ for the Elvis et al. (\cite{elvis}) values.
 A maximum likelihood analysis of the results
yields a mean  $N_{\rm H}$-excess of 
$(0.48\pm0.23)\cdot 10^{20}{\rm cm}^{-2}$ in the XBL sample. In the RBL 
sample the individual errors of the fitted \nh values are generally large
and therefore no statement about deviations from galactic \nh can be 
made. 
Note that due to poor energy resolution in the soft band of the {~PSPC} 
an apparent excess of \nh may also be caused by a steepening of
the intrinsic spectrum.

As for the fainter objects the statistical error of the measured
 \ax is of the same order as the expected variations within the sample, we   
 used the maximum likelihood method to deconvolve the measurement errors
and the intrinsic distribution of the X-ray spectral indices.

We find \am$ = 1.30\pm0.13$, $\sigma_{\rm{G}} = 0.32 \pm 0.12$ for
the RBL sample (34 objects) and  \am$ = 1.40\pm0.09$,
 $\sigma_{\rm{G}} = 0.32 \pm 0.07$ for the XBL sample (40 objects). 

\begin{figure}[htb]
\par\centerline{\psfig{figure=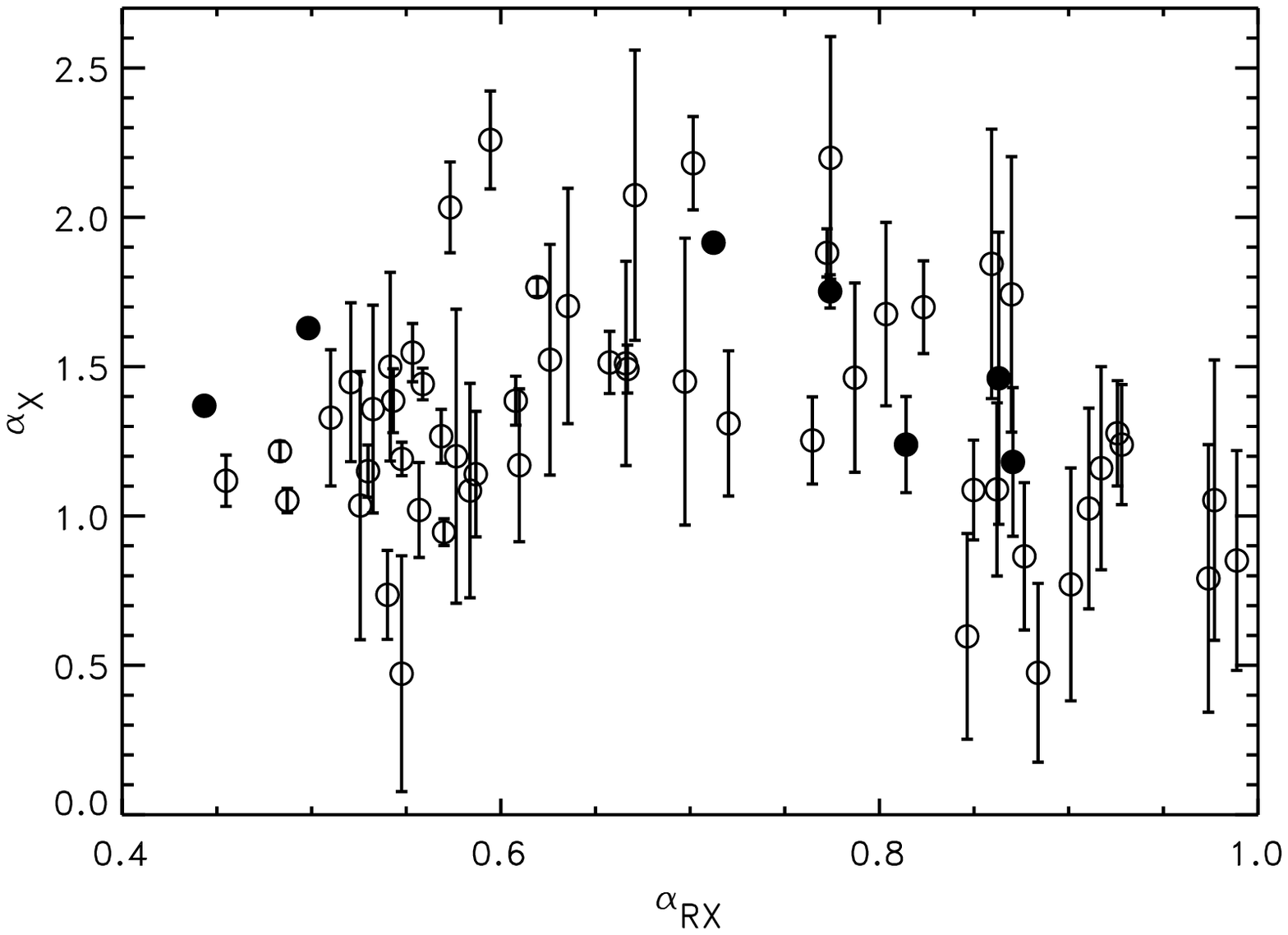,width=8.8 truecm}}
\caption[]{\label{banana}X-ray  spectral index \ax versus broad band index 
\arx, sources with errors $\Delta$\ax $<$ 0.5 only, 
filled circles: EGRET detected objects} 
\end{figure}

Although no significant difference can be found between the mean spectra 
of RBLs and XBLs, there is a dependence of the spectral index on 
the characterising parameter \arx. As can be read from
Fig.~\ref{banana}, objects with extreme values of \arx  on both sides tend to
have flatter X-ray spectra than  intermediate objects. 
For the RBLs (\arx$ > 0.75$) a Spearman Rank test yields an 
{\em anti}correlation of \ax and \arx with 99.9\% probability.
For the XBLs  (\arx$< 0.75$) the significance of the positive 
correlation is 98\%. The steepest X-ray spectra are therefore found
for objects which fall into the gap between XBLs and RBLs 
($0.6<$\arx$<0.8$) for which X-ray spectral indices up to \ax$\sim$2.0
are measured.

In a previous paper (Brunner et al. \cite{brunner}) we stated that 
the mean X-ray spectral index $\langle$\ax$\rangle$
 of RBLs is similar to the mean optical to X-ray  
spectral slope $\langle$\aox$\rangle$, 
whereas the mean X-ray spectrum of {~FSRQs} is 
significantly flatter than their optical to X-ray broad band spectrum.   
However, a relatively large dispersion of the differences 
\aox$-$\ax was found within the RBL sample. 
In the larger sample presented here a dependence of \aox$-$\ax on \arx is
visible (Fig.~\ref{ox_x}). XBLs show a steepening of the spectrum
(\aox$-$\ax$<0$), whereas RBLs can exhibit both steepening or flattening,
depending on \arx. This explains the mean $\langle$\aox$-$\ax$\rangle$
being zero with a large dispersion of the individual values in an RBL 
sample.  The extreme RBLs (\arx$\sim 0.9$) exhibit  spectral flattening
of the same amount as {~FSRQs} (\aox$-$\ax$=0.6$, Brunner et al. \cite{brunner}). 

The spectral results are available as a more detailed version of 
Tab. 3 via WWW (http://astro.uni-tuebingen.de/prepre/), where also
an electronic version of this paper can be found.

\section{Simulated spectra}

The dependence  of the X-ray spectral index on the radio to 
X-ray broad band index \arx can be explained qualitatively 
by a two component spectrum as resulting from synchrotron self-Compton 
jet models (e.g. K\"onigl \cite{koenigl}, Ghisellini et al. \cite{GMT}).
The correlated variations in \arx and \ax are then caused
by a varying high frequency cutoff of the synchrotron component. 

\begin{figure}[htb]
\par\centerline{\psfig{figure=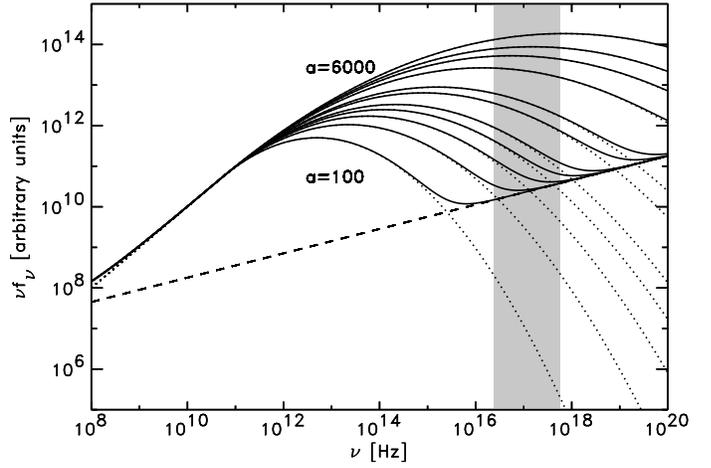,width=8.8 truecm}}
\caption[]{\label{simul}Model spectra used for the simulations. Dotted:
\cs, dashed: \ch, solid: total, shaded area: {~ROSAT} {~PSPC} energy range}
\end{figure}

In order to test this hypothesis we simulated {~ROSAT} {~PSPC} spectra with
a simple two component spectrum as a sum of a parabolically steepening soft 
component \cs and power law hard component \ch (see Fig.~\ref{simul}):
\[ C_{\rm S}(\nu) = 1 \quad \quad  \mbox{ for }\;   \nu<\nu_1 \]
\[ {\rm log}_{10}C_{\rm S}(\nu) = 
\left(\frac{{\rm log}_{10}\nu_1-{\rm log}_{10}\nu}
      {{\rm log}_{10}a}\right)^{\eta}
\quad  \mbox{ for } \; \nu>\nu_1 \]    
\[ C_{\rm H}(\nu) = N \cdot \nu^{-\alpha_{\rm H}} \]
Below $\nu_1$ the soft component \cs  mimics the flat
radio spectrum which BL Lacs have in common; above $\nu_1$ \cs 
is a parabola in the ${\rm log}\; \nu$ -- ${\rm log}\; f_{\nu}$ plane.   
The normalization $N$ of the hard component was chosen so that
\ch(1keV) and \cs(5 GHz) result in a given $\alpha_{\rm rx,max}$.
The extent of the soft component can be varied with the parameter $a$.
Figure \ref{simul} shows the set of calculated spectra using the  
parameters from Tab. \ref{para}.

\begin{table}
\caption[]{\label{para}Parameters of model spectra}
\begin{flushleft}  
\begin{tabular}{cccccc}
\hline\noalign{\smallskip}
$\nu_1$        & $\eta$ & $\alpha_{\rm H}$ & $\alpha_{\rm rx,max}$ & & \\ 
\noalign{\smallskip}
\hline\noalign{\smallskip}
$ 5\cdot10^{10}$& 2.00   & 0.70       & 0.90             & & \\ 
 & & & & \\
\noalign{\smallskip}
\hline\noalign{\smallskip}
\multicolumn{5}{c}{Sequence of variable parameter $a$:} \\ 
\noalign{\smallskip}
\hline\noalign{\smallskip}
100 & 200   &300 & 400 & 500 & 800 \\
1000 & 2000 & 3000 & 4000 & 6000 & \\
\noalign{\smallskip}
\hline    
\end{tabular}
\end{flushleft}
\end{table}

From the resulting spectra corresponding {~ROSAT} {~PSPC} pulse height spectra 
were determined by applying a galactic absorption model (Morrison \& 
McCammon \cite{mmcc}) with $N_{\rm H} = 3\cdot10^{20}{\rm cm}^{-2}$
and folding the spectra with the {~PSPC} efficiency
and detector response matrix. The resulting pulse height spectra were fitted 
with an absorbed power law model in the same way as the BL Lac 
spectra. Broad band spectral indices \arx, \aro, and \aox were determined 
from the flux values  at 5 GHz, 5517\AA, and 1 keV  of each spectrum.

\begin{figure}[htb]
\par\centerline{\psfig{figure=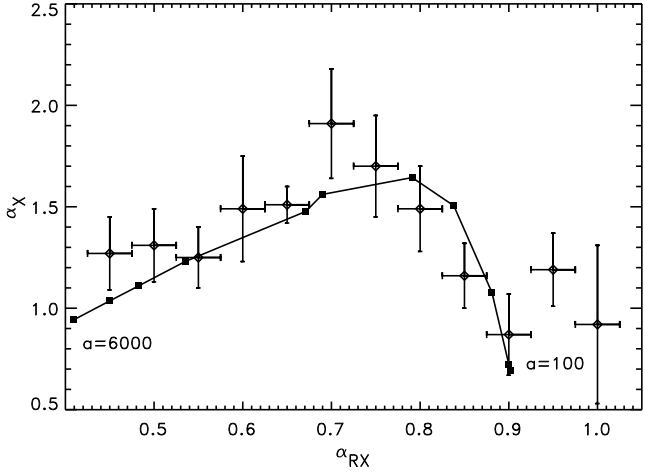,width=8.8 truecm}}
\caption[]{\label{comp}Comparison of simulated (filled squares and solid line)
 and measured (diamonds with error bars) spectra in the \arx -- \ax plane.
The measured spectra were averaged in bins of $\Delta$\arx =0.05.}
\end{figure}

\begin{figure}[htb]
\par\centerline{\psfig{figure=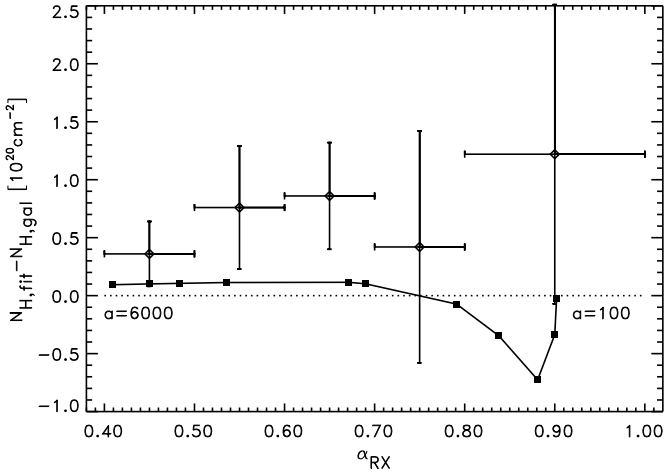,width=8.8 truecm}}
\caption[]{\label{nh}Comparison of simulated (filled squares and solid line) and 
measured (diamonds with error bars) spectra in the \arx -- $\Delta$\nh plane.
The measured $\Delta$\nh values were averaged in the bins indicated by the 
horizontal bars}
\end{figure}

The locations of simulated and observed spectra in the \arx -- \ax plane
are plotted in Fig.~\ref{comp}. 
Mean values of \ax with $1\sigma$ errors have
 been determined in each \arx interval using  maximum 
likelihood contours. We find that the two component model is 
able to 
reproduce the measured interdependence of \arx and \ax.
As curvature of the incident photon spectra is able to cause deviations
of the \nh values resulting from single power law fits, we also compared the
resulting \nh in the simulated and measured spectra.
The deviations in \nh as derived from the simulations are small 
($< 6 \cdot 10^{19}{\rm cm}^{-2}$) and thus are hard to detect in individual
spectra. Averaging the $(N_{\rm H}-N_{\rm H,gal})$ values in bins of \arx using 
the ML  method 
results in an overall excess of the measured \nh values (Fig.~\ref{nh})
over the simulations. Possibly this excess is caused by intrinsic 
absorption in individual sources.

\begin{figure}[htb]
\par\centerline{\psfig{figure=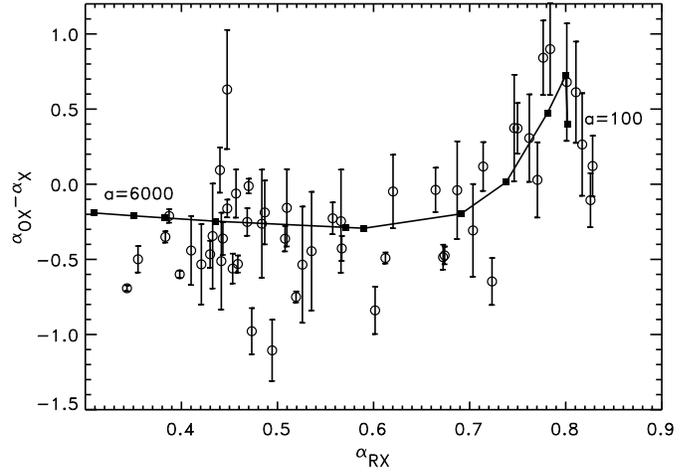,width=8.8 truecm}}
\caption[]{\label{ox_x} Flattening (\aox$-$\ax$>0$) or steepening 
(\aox$-$\ax$<0$)
of the X-ray spectra with respect to optical to X-ray broad band spectra 
vs. \arx. Circles: measured spectra with errors 
$\Delta($\aox$-$\ax$)<0.4$), filled squares and solid line: simulated spectra.}
\end{figure}

The simulations are able to reproduce the measured dependence of the
change in the spectral slope between the optical to X-ray broad band 
spectrum and the X-ray spectrum, (\aox$-$\ax) on \arx (Fig.~\ref{ox_x}). 

Note that the $\alpha_{\rm rx}$ values of the simulated spectra 
cannot exceed $\alpha_{\rm rx,max}$, which was set to 0.9.
Therefore with this parameter set the model does not cover
objects with higher \arx (Figs.~\ref{banana} -- \ref{ox_x}). However, the number
of objects with \arx significantly exceeding 0.9  is small (Fig.~\ref{banana}).

\begin{figure}[htb]
\par\centerline{\psfig{figure=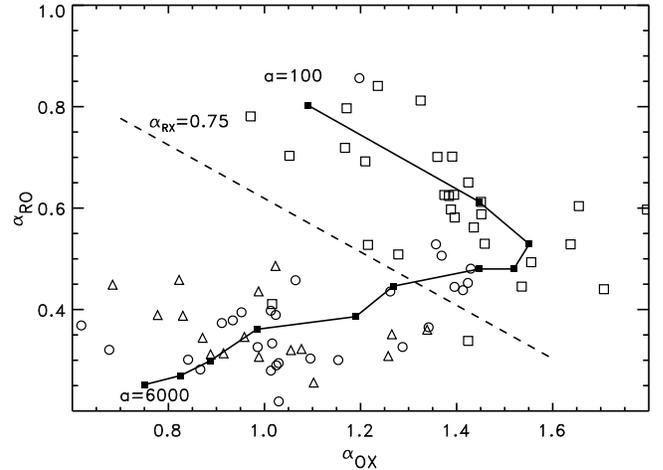,width=8.8 truecm}}
\caption[]{\label{ro} Locations of object and simulated broad band spectra
in the $\alpha_{\rm ox}$ -- $\alpha_{\rm ro}$ plane. 
Triangles: EMSS sample, rectangles: 1~Jy sample, circles: others, 
filled squares and solid line: simulated spectra. 
The division between XBLs and RBLs (\arx=0.75) is 
marked by the dashed line.}
\end{figure}

In order to test whether the two component models do fit the whole radio to 
X-ray continuum, the broad band spectral indices $\alpha_{\rm ox}$ and 
$\alpha_{\rm ro}$ of  simulated and object spectra were compared 
(Fig.~\ref{ro}). The path of the 
model spectra with varying parameter $a$ in the  $\alpha_{\rm ox}$ -- 
$\alpha_{\rm ro}$ plane reasonably well reproduces  the
distribution of the measured broad band spectra.

\section{Inverse Compton beaming factors}

As inverse Compton beaming factors \del are often used to estimate 
the viewing angles of radio sources, we investigated the dependency
of \del on \arx and \ax in our sample.

The Doppler beaming factor $\delta$ in a relativistic jet can be estimated
by the condition that the inverse Compton flux from synchrotron self-Compton 
models must not exceed the observed X-ray flux.
A method for the calculation of \del from radio brightness 
temperature and X-ray flux was given by Marscher (\cite{marscher}):
\begin{equation}   
 \delta_{\rm IC}=
   f(\alpha)F_{\rm m}\left(\frac {{\rm ln}(\nu_{\rm b}/\nu_{\rm m})}
                                  {F_{\rm x}\theta_{\rm d}^{6+4\alpha}
 \nu_{\rm x}^\alpha\nu_{\rm m}^{5+3\alpha}}\right)^{1/(4+2\alpha)}\cdot (1+z) 
\end{equation}
$F_{\rm m}\;$[Jy]: Synchrotron flux at $\nu_{\rm m}$[GHz]\\
$F_{\rm x}\;$[Jy]: X-ray flux at $\nu_{\rm x}$[keV]\\
$\theta_{\rm d}\;$[mas]: VLBI core size\\
$\nu_{\rm b}\;$[Hz]: Synchrotron high frequency cutoff\\  
$\alpha$: Optically thin synchrotron spectral index \\
$f(\alpha)\simeq 0.08\alpha + 0.14$ (Ghisellini et al. \cite{ghisellini})

\begin{figure}[htb]
\par\centerline{\psfig{figure=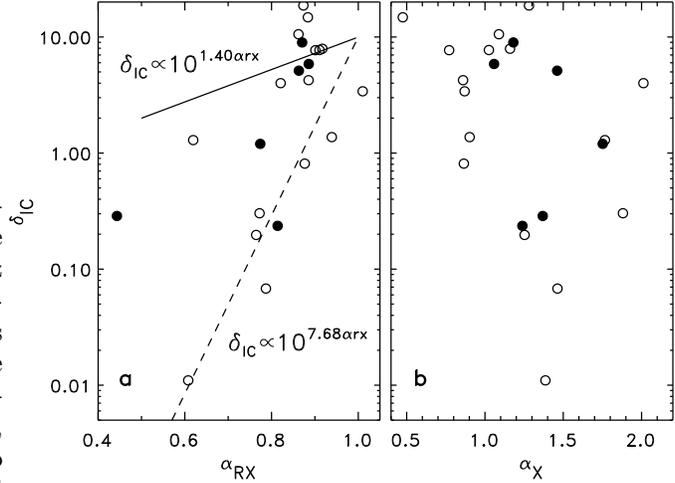,width=8.8 truecm}}
\caption[]{\label{delta}a) Inverse Compton beaming factor $\delta_{\rm IC}$
 vs. \arx. 
 b) $\delta_{\rm IC}$ vs. X-ray spectral index
(filled circles: EGRET detected objects). }

\end{figure}

We calculated Doppler factors \del from the compilation of VLBI data
by Ghisellini et al. (\cite{ghisellini}) and the {~ROSAT} {~PSPC} fluxes at 1 keV;
$\alpha=0.75$ and $\nu_{\rm b}=10^{14}$Hz was assumed.
The results are given in Table 3.
The distributions of the objects
in the \arx -- \del and \ax -- \del planes in Fig.~\ref{delta}
show a strong correlation 
of \del with \arx. 
As the subsample with available VLBI data contains predominantly
RBLs, which show an anticorrelation of \arx and \ax, \del
and \ax are also anticorrelated.
Looking at Eq. (1) two reasons may be responsible for the
correlation of \arx  and \del:

1. The values of \del must be considered as lower limits,
as direct synchrotron emission may contribute to or even dominate the X-ray 
flux and thus the inverse Compton flux $F_{\rm IC}$ is overestimated.
In this case \del will be underestimated by the factor 
\[ \delta_{\rm true}/\delta_{{\rm IC}}=(F_{\rm x}/F_{\rm IC})^{-1/5.5} \] 
Assuming that the diversity in 
\[ \alpha_{\rm rx}=({\rm log}_{10}(F_{\rm 5GHz})
                   -{\rm log}_{10}(F_{\rm 1keV}))/7.68 \]
is caused by a more or less energetic synchrotron component,
\[ \delta_{\rm true}/\delta_{\rm IC} = 10^{1.40\cdot\alpha_{\rm rx}}\]
results. This function is indicated in Fig.~\ref{delta} a) as solid line.

2. Both \arx and \del depend on the radio flux of the source.  
The total fluxes  $F_{\rm 5GHz}$ determining \arx and the VLBI core fluxes 
$F_m$  strongly correlate. Varying the core flux $F_{\rm m}$ and 
$F_{\rm 5GHz}$
by the same factor while leaving $F_{\rm x}$ constant results in  
\[ \delta_{\rm IC} \propto 10^{7.68\cdot \alpha_{\rm rx}} \]     
as indicated by the dashed line in Fig.~\ref{delta} a).

Figure \ref{delta} shows that the first possibility, overestimation
of the Compton flux, cannot fully account for the measured correlation
and viewing angle effects cannot be ruled out to cause the differences
in \del.

Ghisellini et al. (1993) noted that BL Lacs show a broader 
distribution of \del than flat spectrum radio quasars ({~FSRQs})
with a tail towards low values of \del.
It is apparent from Fig.~\ref{delta} that 
one group of objects has properties very similar to flat spectrum radio 
quasars: flat X-ray spectra,  \arx$\sim 0.9$, and $\delta_{\rm IC}=1..10$.
The corresponding properties of {~FSRQs} are:
$\langle\alpha_{\rm x}\rangle = 0.59$, 
$\langle\alpha_{\rm rx}\rangle = 0.88$, (Brunner et al. \cite{brunner} ),
 $\delta_{\rm IC}=1..10$  
(Ghisellini et al. \cite{ghisellini}).
The remaining  RBL  objects have steeper X-ray spectra,
lower \arx\ and lower \del.

\section{EGRET detected objects}

8 objects of the sample have been detected in high energy 
$\gamma$-rays ($>$100 MeV) by EGRET on the Compton
Gamma Ray Observatory: AO~0235+16, PKS~0537-44, S5~0716+71,
S4~0954+65, MARK~421, ON~231, PKS~2005-48 (4--5$\sigma$), and PKS~2155-30 
(von Montigny et al. \cite{montigny}, Vestrand et al. \cite{vest}).  
The EGRET objects are indicated by filled symbols in Figs. 
\ref{banana} and \ref{delta}.
All these objects  except the low distance XBLs MARK~421 and PKS~2155-304 have 
\arx $>$ 0.7. As can be read from Fig.~\ref{banana}, the EGRET sources  
 show the same \arx -- \ax dependence as the remaining 
objects. Note that not only  ``FSRQ-like'' BL Lac objects have been 
detected by EGRET, but also objects with steep X-ray spectra and 
moderate \del.

\section{Comparison with other ROSAT studies}

During the refereeing process of this paper we learned
that a number of {~ROSAT} studies using different BL Lac samples
are going to be published.
In this section we will briefly discuss their results in comparison
with ours. 

Based on an analysis of 12 objects from the 1 Jy sample Comastri et al.
(\cite{comastri}) show that the more
radio bright objects ($\alpha_{\rm rx} > 0.75$) in their 
radio selected sample on average have flatter  X-ray spectra than 
the more X-ray bright ($\alpha_{\rm rx} < 0.75$) ones.
As their sample covers only the radio bright part of the \arx distribution,
this finding fits well to
our overall picture of the spectrum of BL Lacs,
where objects of intermediate \arx do have the steepest
X-ray spectra. 

A detailed investigation of the 
{~ROSAT} observations of the 1 Jy sample has been undertaken by 
Urry et al. (\cite{urry}). 
The above authors both interprete steep X-ray spectra of BL Lacs as 
a sign for synchrotron emission, while flat X-ray spectra should
be dominated by self-Compton emission.
Perlman et al. (\cite{perlman}) performed a {~ROSAT} investigation
of the EMSS XBL sample and found a distribution of X-ray spectra  
similar to the 1~Jy sample.

By considering the whole sample of BL Lac objects we are able to
verify the view, that BL Lacs except the  extreme RBLs are 
dominated by synchrotron emission. We show that the extreme 
XBLs have flat X-ray spectra caused by synchrotron spectra with
cutoff energies beyond the soft X-ray band.  
Our spectral simulations yield a good measure for the synchrotron
cutoff energy and show that the range of  cutoff energies is
large.   

Padovani\&Giommi (\cite{PG96}) have calculated X-ray spectral indices
in a large sample of BL Lacs from hardness ratios provided by the {~ROSAT} 
WGA catalogue (White et al. \cite{WGA}) and with this method 
obtain a similar dependency of \ax on the radio to X-ray flux ratio as we do.

\section{Discussion}

We find that the broad distribution of spectral slopes
in the soft X-ray spectra of BL Lacs is due to a strong
dependence of \ax on the broad band spectral index \arx.
Objects with extreme values of \arx exhibit flat X-ray
spectra, whereas intermediate objects have steeper 
spectra. The symmetry of this dependence prevented the detection
of significant differences between the mean X-ray spectra of XBLs and RBLs
in previous investigations (Worrall \& Wilkes \cite{ww}, 
Sambruna et al. \cite{sambruna}, Lamer et al.
\cite{lamer}). 

By comparison with simulated {~PSPC} spectra we showed
that a two component model with a hard power law ($\alpha=0.7$)
component and a steepening soft component is appropriate
to explain the observed spectra.
The frequency where the components intersect each other is below
the soft X-ray band for extreme RBLs and crosses the energy band of the 
{~ROSAT} {~PSPC} with declining \arx.
In the framework of the SSC models this means that Compton emission
causes the flat X-ray spectra of extreme RBLs,
whereas the likewise flat X-ray spectra of extreme XBLs are due to synchrotron 
emission. The steep X-ray spectra of objects with intermediate spectral
energy distribution ($0.6<$\arx$<0.8$) represent the first direct 
evidence of the synchrotron high energy cutoff.

Padovani \& Giommi (\cite{PG95a}) explain the different spectral energy 
distributions of XBLs and RBLs by 
different energy cutoffs of the synchrotron spectra. 
They postulate the cutoff energy being intrinsic properties of the sources
without discussing the physics of the emission processes. 
This scenario also is the most straightforward explanation
for our findings, including the correlation of \arx and \ax for the XBL
subsample. 
The wide range in synchrotron cutoff energies, and consequently
the cutoff in  the energy spectrum of the relativistic electrons,
has to be explained.  
The SSC cooling of the jet electrons may be more efficient in the more 
powerful jets of {~FSRQs} and RBLs than in the jets of XBLs.
Ghisellini \& Maraschi (\cite{GM94}) proposed a more rapid
cooling of jet electrons in {~FSRQs} by external UV photons compared to
jets of BL Lac objects. It is conceivable that RBLs are intermediate 
objects between XBLs and {~FSRQs} regarding the ambient photon density.

The more physically motivated beaming models,
such as the ``accelerating jet'' model (Ghisellini \& Maraschi 
\cite{GM}) and the ``wide jet'' model (Celotti et al. \cite{celotti}) do
not as naturally satisfy our data. The crossover frequencies of soft 
and hard components in the spectra calculated by Ghisellini and Maraschi
(\cite{GM}) do not span a sufficient range and do not move across the soft 
X-ray range when tilting the viewing angle, as required by the new data.
Nevertheless, further tuning of the free parameters may provide spectra 
which are in accordance with the measurements. The {~ROSAT} spectra 
therefore are suitable to constrain the beaming models.

\begin{acknowledgements}
This work was supported by DARA under grant 50 OR 90099
and has made use of the NASA/IPAC Extragalactic Database (NED)
which is operated by the Jet Propulsion Laboratory, Caltech, under
contract with the National Aeronautics and Space Administration.
\end{acknowledgements}

\end{document}